\documentclass[a4paper]{panl}
\usepackage{cite}
\usepackage{wrapfig}
\usepackage{graphicx}
\usepackage{amssymb}
\usepackage{amsfonts}
\usepackage{amsmath}
\usepackage{longtable}
\usepackage{rotating}
\usepackage{lscape}
\usepackage{epsfig}
\usepackage{multirow}

\usepackage{hyperref}
\usepackage{xcolor}

\originalTeX
\begin{document}
% Journal sections (see http://pkp.jinr.ru/index.php/PEPAN_LETTERS/about/editorialPolicies#focusAndScope)
%\issuearea{Physics of Elementary Particles and Atomic Nuclei. Theory}
% or in Russian
%\issuearea{ФИЗИКА ЭЛЕМЕНТАРНЫХ ЧАСТИЦ И АТОМНОГО ЯДРА. ТЕОРИЯ}

\title{On the Principle of Relativity of Inertia in both General and Entangled Relativities}% \\ О принципе относительности инерции как в общей, так и в запутанной теории относительности}
\maketitle
\authors{O.\,Minazzoli$^{a,b,}$\footnote{E-mail: olivier.minazzoli@oca.eu}}%,
%S.\,Author$^{b,}$\footnote{E-mail: second.author@email.ru}}
\setcounter{footnote}{0}
%\authors{И.О.\,Первыйавтор$^{a,b,}$\footnote{E-mail: first.author@email.ru(русский вариант)},
%И.О.\,Второйавтор$^{b,}$\footnote{E-mail: second.author@email.ru(русский вариант)}}
\from{$^{a}$\,Universit\'e C\^ote d'Azur, Observatoire de la C\^ote d'Azur, CNRS, Artemis, bd. de l'Observatoire, 06304, Nice, France.}
%\from{$^{a}$\,Место работы автора 1}
\from{$^{b}$\,Bureau des Affaires Spatiales, 2 rue du Gabian, 98000  Monaco.}
%\from{$^{b}$\,Место работы автора 2}

\begin{abstract}
% Russian translation of the abstract
%Аннотация на русском языке. \
\vspace{0.2cm}
Entangled Relativity is a novel theory of relativity that offers a more economical approach than General Relativity. It successfully recovers both General Relativity and standard quantum field theory within a specific (yet generic) limit. Furthermore, Entangled Relativity precludes the existence of spacetime devoid of the matter that permeates it. Consequently, I argue that Entangled Relativity is not only preferable from the standpoint of Occam's razor, due to its economical nature, but it also aligns more closely with Einstein's original vision for a satisfactory theory of relativity.
%Entangled Relativity is a novel general theory of relativity that is more economical than General Relativity, while it recovers both General Relativity and standard quantum field theory in a specific (but generic) limit. Besides, Entangled Relativity prevents the existence of spacetime without matter that permeats it. Therefore, I argue that not only Entangled Relativity is preferable from the economical standpoint of Occkham's razor, but that it also adheres more closely to Einstein's original idea of a satisfying theory of relativity.
\end{abstract}
\vspace*{6pt}

\noindent
PACS: 01.65.+g; 04.50.Kd; 04.60.Gw

%--------------------------------------------------------------
\label{sec:intro}
\section{Introduction}

The \textit{principle of the relativity of inertia} is one of the three foundational principles of Einstein's General Theory of Relativity \cite{einstein:1918an}. However, Einstein later acknowledged that General Relativity does not fully satisfy this principle \cite{hoefer:1995cf,pais:1982bk}. This principle asserts that spacetime should be completely determined by matter, rather than being only partially influenced by it. Notably, it implies that spacetime cannot exist without the matter that generates it \cite{einstein:1918an,einstein:1918sp}. In other words, according to this principle, any theory allowing for the existence of classical vacuum solutions in spacetime should be considered invalid. Yet, General Relativity permits purely vacuum solutions, such as Minkowski, Schwarzschild, or Kerr spacetimes when the cosmological constant is set to zero, and de Sitter or Schwarzschild-de Sitter spacetimes when it is not.

In this proceeding, after revisiting the issue within General Relativity, both in its historical context and in modern physics, I will present Entangled Relativity. I will explain why it aligns more closely with this principle and why it offers a more parsimonious formulation compared to General Relativity.
%
%is one of the three founding principle of Einstein's general theory of relativity (General Relativity) \cite{einstein:1918an}, although Einstein later aknowledged that General Relativity does not satisfy it \cite{hoefer:1995cf,pais:1982bk}. This principle demands that spacetime is completely determined by matter instead of being partially determined by it. The principle notably implies that spacetime cannot exist without matter that generates it  \cite{einstein:1918an,einstein:1918sp}. In other words, a theory for which spacetime solutions in a vacuum exist, in the classical sense, should be prohibited according to this principle. However, General Relativity allows for purely vacuum solutions such as Minkowski, Schwarzschild or Kerr, when the cosmological constant is set to zero, and de Sitter or Schwarzschild-de Sitter when it is not.
%In this proceeding, after reminding what the problem is in General Relativity, both in the historical context and in the context of modern physics, I shall present Entangled Relativity and why it (better) aligns with this principle, in addition to being more parsimonious than General Relativity in its formulation. 

%--------------------------------------------------------------
\section{The relativity of inertia in General Relativity}

The principle of relativity of inertia,\footnote{First mentioned in \cite{einstein:1913br} as ‘the hypothesis of the relativity of inertia’.} which Einstein also named \textit{Mach's principle} in \cite{einstein:1918an}, stems from the metaphysical assertion that motion can only be defined with respect to matter, rather than from space (or spacetime) itself, \textit{\`a la Newton}. In other words, inertia is relative to surrounding matter fields, and not to an underlying absolute structure. As Norton writes, ``The doctrine of the relativity of motion is attractive for its simplicity. According to it, the assertion that a body moves can mean nothing more than that it moves with respect to other bodies'' \cite{norton:1995cf}. The endeavor to satisfy this principle reflects Einstein's engagement in a longstanding debate which, in some aspects, dates back to antiquity \cite{barbour:1995cf}, but gained prominence with Newton’s theory of gravitation. Newton proposed that \textit{dynamics} should be defined with respect to an \textit{absolute} space, which provides the framework for phenomena to occur. Conversely, Huygens and Leibniz argued that motion is \textit{relative} between bodies, as opposed to being relative to an absolute structure. Mach revisited this debate, advocating for the relativity of motions, i.e., the relativity of inertia. Deeply influenced by Mach,\footnote{However, there is debate over whether Einstein fully grasped Mach's philosophy and whether Mach and his pupil, Petzoldt, truly endorsed Einstein's theory as genuinely \textit{Machian} \cite{borzeszkowski:1995cf,krauss:2023cf}. (From my perspective, what is most relevant is Einstein’s intuition, particularly during the period leading to General Relativity, rather than its alignment with the work of some contemporaries).} Einstein thus proposed that a satisfactory theory of relativity should adhere to a principle of relativity of inertia. He writes:
``The [\textit{Entwurf}] theory sketched here overcomes an epistemological defect that attaches not only to the original theory of relativity, but also to Galilean mechanics, and that was especially stressed by E. Mach. It is obvious that one cannot ascribe an absolute meaning to the concept of acceleration of a material point, no more so than one can ascribe it to the concept of velocity. Acceleration can only be defined as relative acceleration of a point with respect to other bodies.'' \cite{einstein:1914sr}.

After finalizing the equations of General Relativity (including a cosmological constant), Einstein’s definition of the principle of the relativity of inertia is articulated in \cite{einstein:1918an} as:
%After settling on the final forms of the equation of General Relativity (with a cosmological constant), Einstein's definition of the principle of the relativity of inertia becomes in \cite{einstein:1918an}:  
``Mach's principle. The G-field is completely determined by the masses of the bodies', and Einstein completes further down in his manuscript by writing ``With [this principle], according to the field equations of gravitation, there can be no G-field without matter. Obviously, [this principle] is closely connected to the spacetime structure of the world as a whole, because all masses in the universe will partake in the generation of the G-field''. Although his understanding of what a principle of relativity of inertia could mean within the relativistic framework evolved with time \cite{hoefer:1995cf}---with some confusion prior to 1915 between the \textit{principle of relativity} (\textit{covariance principle}) and the \textit{equivalence principle}---
%---with some notable confusion priori to 1915 with the \textit{principle of relativity} (\textit{covariance principle}) and the \textit{equivalence principle}---
the fundamental necessity for this principle, evidently, is that spacetime cannot exist without matter \cite{hoefer:1995cf}. In Einstein’s own words: 
%the underlying aboslute necessity for this principle to be realized is, obviously, that spacetime cannot exist without matter \cite{hoefer:1995cf}. In Einstein's own words: 
"[In] my opinion, the general theory of relativity is a satisfying system only if it shows that the physical qualities of space are completely determined by matter alone. Therefore,  no $g_{\mu \nu}$-field must exist without matter that generates it.'' \cite{einstein:1918sp}---see below for the context of this quote.

%Indeed, in a relativistic theory, inertia is defined from the spacetime metric. Therefore, if a relativistic theory allows for vacuum spacetime solution, it means that inertia could, de facto, be defined without any matter field. This would threfore contradict Einstein's original intention to \textit{relativize} the motions of bodies with each other bodies instead of with respect to an absolute spacetime structure.
Indeed, in a relativistic theory, inertia is defined by the spacetime metric. Therefore, if a relativistic theory permits vacuum spacetime solutions, it implies that inertia could, de facto, be defined without any matter field. This would therefore contradict Einstein's original intention to \textit{relativize} the motions of bodies in relation to each other, rather than with respect to an absolute spacetime structure.

%--------------------------------------------------------------
\subsection{The cosmological constant:~}

While reading Einstein's cosmological paper in isolation might lead one to conclude that he introduced the cosmological constant to enable a static universe, as is often suggested nowadays, placing the paper in the context of his other works reveals that Einstein's true intent in adding a cosmological constant was to satisfy the relativity of inertia. In his cosmological paper, he asserts: 
%While if one reads Einstein's cosmological paper as a standalone paper, one may arrive at the conclusion that Einstein introduced the cosmological constant in order to have a static uniserve, as it is often presented these days, putting the paper in the context of all his other papers allows one to understand that what Einstein really wanted by adding a cosmological constant was to satisfy the relativity of inertia. Indeed, in his cosmological paper, he states: 
``[In] a consistent theory of relativity there cannot be inertia relatively to ``space'' but only an inertia of masses relatively to one another'' \cite{einstein:1917co}. 
As Hoefer explains \cite{hoefer:1995cf}, at that time, Einstein believed this requirement would impose two strong, albeit distinct, constraints on the theory and its acceptable solutions:
%As explained by \cite{hoefer:1995cf}, at that time, Einstein thought that this demand would imply two strong, yet very different in nature, constraints on the theory itself and on its potential acceptable solutions. 
\begin{enumerate}
\item The universe should be finite and closed, to avoid boundary conditions without matter, thereby upholding Mach's principle. This constraint serves as a criterion for selecting acceptable solutions among many that do not fulfill this requirement.
%The universe should be finite and closed, in order not to have boundary conditions without matter (which would therefore contradict Mach's principle). Let us stress that this constraint gives a selection criterium of acceptable solutions, among potentially many solutions that do not satisfy this property.
\item The theory should not permit solutions without matter.\footnote{Due to the page limit, I will not address the issue of vacuum solutions with singularities here.} Unlike the first constraint, this is not about selecting potential solutions from a given theory, but a condition for the theory itself.
%The theory should not have matter-free solutions.\footnote{Due to the number of pages limit, I leave aside the issue of the status of allowing for vacuum solutions with singularities in this discussion.} This, on the opposite of the constraint 1., is not a selection of potential acceptable solutions of a given theory, but a condition on the acceptable theory itself.
\end{enumerate}
The cosmological constant seemed to be an elegant mathematical solution that was able to satisify both constaints at a single stroke. Indeed, Einstein's cosmological solution in \cite{einstein:1917co} is constructed in order to respect the first constraint, and it is clear that the cosmological constant is what enables that. 
However, although not explicitly stated in his papers, Einstein was aware that Minkowski spacetime is a trivial vacuum solution of General Relativity without a cosmological constant. For some reason, he also believed that adding a cosmological constant would preclude any vacuum solutions. This is evident from his response to de Sitter's vacuum solution of General Relativity with a cosmological constant \cite{einstein:1918sp}:
%But, although not explicitely written in his papers, Einstein knew that Minkowski was a trivial vacuum solution of General Relativity without a cosmological constant, and for some reasons, he also beleived that adding a cosmological constant to the equation of General Realtivity would prevent the existence of any vacuum solution. This is clear from his response to de Sitter's solution of General Relativity in vacuum but with a cosmological constant \cite{einstein:1918sp}: 
``If the De Sitter solution were valid everywhere, it would show that the introduction of the $\lambda$-term does not fulfill the purpose I intended. Because, in my opinion, the general theory of relativity is a satisfying system only if it shows that the physical qualities of space are completely determined by matter alone. Therefore, no $g_{\mu \nu}$-field must exist (that is, no spacetime continuum is possible) without matter that generates it.''

Following Hoefer in \cite{hoefer:1995cf}, I believe that it is Constraint 2, rather than Constraint 1, that truly encapsulates Mach's principle of Einstein. Hoefer writes: 
%Following  Hoefer in \cite{hoefer:1995cf}, I beleive that it is the constraint 2., rather that the constraint 1., that really captures Mach's principle of Einstein. Hoefer writes: 
``The widespread assumption that a closed, matter-filled cosmology such as Einstein's spherical cosmology must satisfy Mach's Principle is questionable. It is based on the reasoning discussed above, that since in anti-Machian models the trouble seems to come from the boundary conditions, if one eliminates the boundary region one eliminates the problem. But this reasoning is clearly fallacious. There is a missing premise: The only way a model can fail to be Machian is to have an empty boundary region in which an absolute spatiotemporal structure is posited. [...] In the meantime, it seems more clear that we can (as did Einstein) rule out some models of GTR as definitely anti-Machian and use these judgments as constraints on any explication of Mach's Principle. The clearest case is that of empty spacetimes. Since they do not contain any matter-energy and have a definite spatial (and hence inertial) structure, they clearly run contrary to the core of Mach's ideas on the origin of inertia. Therefore, I believe that Einstein was absolutely right to demand [2.] as a necessary condition for the relativization of inertia, or satisfaction of Mach's principle by a gravitation theory. Demand [2.] should remain our most secure touchstone in theorizing about how to create a Machian gravitation theory.'' Indeed, as we will see below, our modern understanding of physics through the prism of quantum field theory imply that matter-free parts of the universe do not exist, even at an hypothetical boundary.

%--------------------------------------------------------------
\subsection{Our universe:~}

If one examines our universe, there does not appear to be any conflict with respect to Mach's principle: matter exists everywhere, and local inertial structures are determined by embedding any local spacetime into more global ones. These global structures are ultimately determined by matter, up to the homogeneous and isotropic zeroth-order model corresponding to the Friedmann-Lemaître-Robertson-Walker metric. Therefore, if one considers our universe as one of the many potential solutions of General Relativity, this specific \textit{realization} seems clearly acceptable in terms of the principle of the relativity of inertia. Nevertheless, I will argue that there is still an issue with respect to the principle of relativity of inertia in General Relativity.
%If one looks at our universe, there really does not seem to be any problem with respect ot Mach's principle: matter exist everywhere, and local inertial structure are determined by embeding any local spacetime onto more global ones, which are determined by matter, utlimately up to the homogeneous and isotropic zeroth order model corresponding to the Friedmann-Lemaître-Robertson-Walker metric. So, if one considers our universe to be one of the many potential solutions of General Relativity, this specific \textit{realization} clearly seems acceptable in terms of the principle of the relativity of inertia. Nevertheless, I will argue that there still is an issue with respect to the principle of relativity of inertia in General Relativity.

%--------------------------------------------------------------
\subsection{Dual ontologies, the core of the problem:~}

Although the General Relativity description of our universe seems to satisfy the principle of relativity of inertia, the issue with General Relativity is that inertia can, in principle, be either a relativistic property or an intrinsic (or absolute) one. This dual possibility is not internally consistent. An example can be seen with Kerr black holes. A Kerr black hole is a vacuum solution of General Relativity. This mathematical solution represents an entire universe that is asymptotically flat and contains a rotating black hole at the center of the coordinate system. A Kerr black hole is fundamentally different from a non-rotating Schwarzschild black hole; notably, one cannot transform one into the other by a change in the coordinate system, such as adopting a rotating frame. However, since a Kerr black hole is a vacuum solution, it implies that rotation in this context is an intrinsic (or absolute) property of the black hole, rather than a relative one. This suggests that inertia can be an absolute structure in General Relativity, while it can also be a relative one (as exemplified by the solution corresponding to our actual universe). This dual ontology of inertia in General Relativity is logically inconsistent. In the concluding discussion of his book \textit{The Meaning of Relativity} \cite{einstein:1921bk}, Einstein states: 
%Although in the General Relativity's description of our universe, the principle of relativity of inertia seems to be satisfied, the issue with General Relativity is that inertia can, in principle, be both a relativist property, or an instrinsic (or absolute) one. In terms of internal logic, this is not consistent. This can be exemplified with Kerr black-holes. A Kerr black-hole is a vacuum solution of General Relativity. The mathematical solution represents an entire universe that is assymptotically flat, and which contains a rotating black-hole at the center of the coordinate system. A Kerr black-hole is genuinely different from a non-rotating Schwarzschild black-hole: notably, one cannot transform one into the other by a change of the coordinate system, for instance with a coordinate system that would be rotating as well. But because a Kerr black-hole is a vacuum solution, it means that a rotation in this case is an intrinsic (or absolute) property of a black-hole, rather than a relative one.  This means that inertia can be an aboslute structure in General Relativity, while it can also be a relative one (as exemplified by the solution corresponsing ot our actual universe). This dual ontology of inertia in General Relativity is logically inconsistent. In the concluding discussion of his book \textit{The Meaning of Relativity} \cite{einstein:1921bk}, Einstein writes: 
``As it is an unsatisfactory assumption to make that inertia depends in part upon mutual actions, and in part upon an independent property of space, Mach's idea gains in probability.''

%--------------------------------------------------------------
\subsection{A modern physics' take on the problem:~}

According to quantum field theory, classical physics simply corresponds to paths with stationary quantum phases in the path integral over all non-redundant possible paths. The reason is that for large actions relative to Planck's quantum of action $\hbar$, paths far from those producing a stationary phase interact destructively with each other. In contrast, there is constructive interference for paths with stationary phases and those close enough. The Core theory of physics\footnote{That is, the current standard model of physics, as named by Wilczek \cite{wilczek:2016bk}.} is expressed as follows:
%According to quantum field theory, classical physics simply corresponds to paths with stationnary quantum phases in the path integral over all non-redundant possible paths. The reason being that for large actions with respect to Planck's quantum of action $\hbar$, paths that are far from the ones that produce a stationnary phase interact destructively between each other, whereas there is a constructive interference for stationnary paths, and paths that are close-by enough. The Core theory of physics\footnote{That is, the current standard model of physics, as named by Wilczek \cite{wilczek:2016bk}.} writes as follows
\begin{equation}
Z_{\textrm{C}} = \int [D g] \prod_i [D f_i] \exp \left[\frac{i}{\hbar c} \int d^4_g x \left(\frac{R(g)}{2 \kappa_{GR}} + \mathcal{L}_m^{SM}(f,g) \right)\right], \label{eq:CorePI}
\end{equation}
where $\int [D]$ relates to the sum over all possible (non-redundant) field configurations,\footnote{Usually, a cut-off scale for the integral is defined, beyond which this formulation is not believed to hold true.} $R$ is the usual Ricci scalar that is constructed upon the metric tensor $g$, $ \mathrm{d}^4_g x := \sqrt{-|g|}  \mathrm{d}^4 x$ is the spacetime volume element, with $|g|$ the metric $g$ determinant.  $\mathcal{L}_m$ is the Lagrangian density of matter, where $f_i$ are the matter fields of the standard model (SM) of particle physics---such as fermions and gauge bosons, and the Higgs. It also depends on the metric tensor, a priori through to the \textit{comma-goes-to-semicolon rule} \cite{MTW}.
Let us note that there are three universal constants in this formulation: the quantum constant $\hbar$ (Planck's), the causal structure constant $c$ and the constant of gravity $G = c^4 \kappa_{GR} / (8\pi)$ (Newton's).

Due to the very nature of Eq. (\ref{eq:CorePI}), solutions devoid of matter fields\footnote{That is, vacuum solutions in the \textit{classical} sense.} cannot exist as long as $\mathcal{L}_m \neq \emptyset$. Consequently, quantum field theory leans towards satisfying the principle of the relativity of inertia, as it implies the presence of matter fields throughout space and time. Notably, this means that matter fields must exist at any boundary, should such a boundary exist, rendering demand 1, mentioned above---which was designed to avert the possibility of \textit{absolute} boundary conditions---entirely unnecessary in this context.
%Because of the very nature of Eq. (\ref{eq:CorePI}), solutions that would be void of matter fields\footnote{That are, vacuum solutions in the \textit{classical} sense.} cannot exist as soon as $\mathcal{L}_m \neq \emptyset$ in Eq. (\ref{eq:CorePI}). Therefore, quantum field theory goes toward a satisfaction of the principle of the relativity of inertia, since it enforces the existence of matter fields everywhere in space and time. Notably, it means that there exist matter fields at any boundary, if such a boundary exists, such that the demand 1. mentionned above---which was designed to prevent the possibility of \textit{absolute} boudary conditions---becomes totally unecessary in this framework.

Nonetheless, two potential questions remain: A/ Why is $\mathcal{L}_m \neq \emptyset$ instead of $\mathcal{L}_m = \emptyset$? B/ Is it still possible, at least theoretically, to have an explicit violation of the relativity of inertia even with matter fields present everywhere? While the first question appears somewhat metaphysical in the context of Eq. (\ref{eq:CorePI}), we will see below that it receives an answer within the framework of Entangled Relativity: specifically, Entangled Relativity precludes having $\mathcal{L}_m = \emptyset$. 
%It nervetheless remains two potential questions: A/ Why $\mathcal{L}_m \neq \emptyset$ rather than $\mathcal{L}_m = \emptyset$? B/ Is it still possible, at least in principle, to have an explicit violation of the relativity of inertia (even) with matter fields everywhere? While the first question seems rather metaphysical in the context of Eq. (\ref{eq:CorePI}), we will see below that it gets an answer in the context of Entangled Relativity: namely, Entangled Relativity forbids to have $\mathcal{L}_m = \emptyset$ in the first place. 
The second question, on the other hand, cannot be resolved without either proving that such solutions are impossible or finding an explicit example of such a violation---like, in the context of General Relativity, G\"odel's famous rotating universe \cite{godel:1949rm}, although this is subject to debate.

%first finding an explicit example with such a violation---such as, although this is up to debate, G\"odel's famous rotating universe in General Relativity for instance \cite{godel:1949rm}---or proving that such solutions cannot exist otherwise. 
%Fortunately, such an example exists, and has been found by G\"odel \com{[citation]}.
%
%%--------------------------------------------------------------
%\subsection{G\"odel's solutions of General Relativity with a cosmological constant:~}

%--------------------------------------------------------------
\section{The relativity of inertia in Entangled Relativity}

As already mentioned, Entangled Relativity precludes the existence of spacetime without matter fields permeating it, while simultaneously being more parsimonious than General Relativity in its formulation. This is due to the non-linear coupling between matter and curvature in the theory's formulation. Entangled Relativity is defined by its path integral, expressed as:
%As mentionned already above, Entangled Relativity prevents the existence of spacetime without matter fields that permeat it, while at the same time it is more parsimonious than General Relativity in its formulation. Both of these properties are due to the non-linear coupling between matter and curvature in the formulation of the theory. The definition of Entangled Relativity is based on its path integral formulation and reads
\begin{equation}
Z_{ER} = \int [D g]  \prod_i [D f_i] \exp \left(-\frac{i}{2 \epsilon^2} \int d^4_g x \frac{\mathcal{L}^2_m(f,g)}{R(g)} \right), \label{eq:ERPI}
\end{equation}
where $\mathcal{L}_m$ is the Lagrangian density of matter fields $f$---which could be the current \textit{standard model of particle physics} Lagrangian density, or most likely a completion of it. Indeed, $Z_{ER}$ cannot even be defined if $\mathcal{L}_m = \emptyset$. In some sense, Entangled Relativity satisfies the principle of relativity of inertia simply because \textit{nothing} is not an option in this framework; whereas pure quantum General Relativity\footnote{That is, assuming $\mathcal{L}_m = \emptyset$ from the start.} can be defined and studied---as one can notably check within the framework of \textit{Causal Dynamical Triangulation} for instance \cite{loll:2008cq}.

Furthermore, the formulation of this theory requires only two universal dimensionful constants instead of three: the causal structure constant $c$, which is implicit in the definition of the four-volume element $d^4_g x$, and a squared quantum of energy $\epsilon^2$. Notably, Planck's quantum of action $\hbar$ does not appear in the formulation, suggesting it is not a fundamental constant in this theory, nor is Newton's constant $G$. Consequently, the Planck length and time are not fundamental in Entangled Relativity, whereas the Planck energy still is.
%Moreover, the formulation of the theory requires only two universal dimensionful constants instead of three: the causal structure constant $c$, which is hidden in the definition of the four-volume element $d^4_g x$, and a quantum of energy squarred $\epsilon^2$. In particular, Planck's quantum of action $\hbar$ does not appear in the formulation of the theory, indicating that it is not a fundamental constant in this theory, nor is Newton's constant $G$. It notably follows that the Planck length and time are not fundamental in Entangled Relativity, whereas the Planck energy still is. 
Indeed, in order to recover standard quantum field theory when gravity is neglected, one can show that $\epsilon$ has to be the reduced Planck energy \cite{minazzoli:2022ar,minazzoli:2023ar}.

%--------------------------------------------------------------
\subsection{Classical physics:~}

While both the perturbative and the non-perturbative quantum behavior of Eq. (\ref{eq:ERPI}) remain to be studied, one can check that the classical field equations that follow from it seem to be consistent with our world---at least, up to further scrutiny. Classical physics corresponds to stationnary phases in Eq. (\ref{eq:ERPI}), leading to the following equations \cite{ludwig:2015pl}:
\begin{equation} \label{eq:metricfe}
G_{\mu \nu} = \kappa T_{\mu \nu} + f_R^{-1} \left[\nabla_\mu \nabla_\nu - g_{\mu \nu} \Box \right] f_R,
\end{equation}
where 
\begin{equation}
T_{\mu \nu} := -\frac{2}{\sqrt{-g}} \frac{\delta\left(\sqrt{-g} \mathcal{L}_{m}\right)}{\delta g^{\mu \nu}}.
\end{equation}
and $G_{\mu \nu}  := R_{\mu \nu} - 1/2 R g_{\mu \nu} $ is the usual Einstein tensor, with
\begin{eqnarray}
\kappa = - \frac{R}{\mathcal{L}_m},\qquad
%f_R = \frac{1}{2} \frac{\mathcal{L}_m^2}{R^2} = \frac{1}{2 \kappa^2},
\left(f := -\frac{1}{2 \epsilon^2}\frac{\mathcal{L}_m^2}{R}, \qquad f_R := \frac{\partial f}{\partial R} = \frac{1}{2 \epsilon^2} \frac{\mathcal{L}_m^2}{R^2} = \frac{1}{2 \epsilon^2 \kappa^2}\right).\label{eq:f_Rkappa}
\end{eqnarray}
Let us note that $\kappa_{GR} = -R/T$ in General Relativity instead.
The stress-energy tensor is not conserved in general, as one has
\begin{equation}
\nabla_{\sigma}\left(\frac{\mathcal{L}_{m}}{R} T^{\alpha \sigma}\right)=\mathcal{L}_{m} \nabla^{\alpha}\left(\frac{\mathcal{L}_{m}}{R}\right). \label{eq:noconsfR}
\end{equation}
The matter field equation, for any tensorial matter field $\chi$, gets modified due to the non-linear coupling between matter and curvature as follows
\begin{equation}
\frac{\partial \mathcal{L}_{m}}{\partial \chi}-\frac{1}{\sqrt{-|g|}} \partial_{\sigma}\left(\frac{\partial \sqrt{-|g|} \mathcal{L}_{m}}{\partial\left(\partial_{\sigma} \chi\right)}\right)=\frac{\partial \mathcal{L}_{m}}{\partial\left(\partial_{\sigma} \chi\right)} \frac{R}{\mathcal{L}_{m}} \partial_{\sigma}\left(\frac{\mathcal{L}_{m}}{R}\right). \label{eq:ERmatter}
\end{equation}
It has been shown that these equations lead to a classical phenomenology that is very similar to, or even indistinguishable from, that of General Relativity in many cases. Surprisingly, they also encompass standard quantum field theory as a limit. This can be traced back to the \textit{intrinsic decoupling} originally discovered in scalar-tensor theories \cite{minazzoli:2013pr}. Indeed, as is typical in $f(R)$ theories, the trace of the metric field equation produces the differential equation for the extra scalar degree of freedom $\kappa = - R/\mathcal{L}_{m}$, which is
%It has been shown that these equations lead to a classical phenomenology that is very close (or even indistinguishable) to the one of general relativity in many cases, while it also (supprisingly) have standard quantum field theory as a limit. It all boils down to the \textit{intrinsic decoupling} that was originally found for scalar-tensor theories in \cite{minazzoli:2013pr}. Indeed, as usual in $f(R)$ theories, the trace of the metric field equation produces the equation for the extra scalar degree-of-freedom $\kappa = - R/\mathcal{L}_{m}$, which is
\begin{equation}
%3 \kappa^{-2}\square \kappa^2=\kappa \left(T-\mathcal{L}_{m}\right). \label{eq:sceq}
3 \kappa^{2}\Box \kappa^{-2}=\kappa \left(T-\mathcal{L}_{m}\right). \label{eq:sceq}
\end{equation}
Therefore, whenever $\mathcal{L}_m = T$ on-shell, the extra degree-of-freedom is not sourced and become constant in many occurrences, and one recovers General Relativity minimally coupled to matter, and without a cosmological constant, to a very good accuracy. Let us recall that $\mathcal{L}_m = T$ for a universe that would entirely be made of dust and electromagnetic radiation for instance, which turns out to be a very good approximation of the current content of our universe. As a side note, whenever $\mathcal{L}_m = T$, one also recovers $\kappa = - R/T$ of General Relativity.

%--------------------------------------------------------------
\subsection{Recovering standard quantum field theory:~}

Interrestingly, all the previous field equations, including those of matter fields, can be recovered by the alternative Einstein-dilaton phase that follows, provided that one has $\mathcal{L}_m \neq 0$:
\begin{equation}
\label{eq:actphi0}
\Theta = \frac{1}{\epsilon^2} \int \mathrm{d}_g^{4} x  ~\frac{1}{\kappa}\left(\frac{R}{2 \kappa} + \mathcal{L}_m \right),
\end{equation}
where $\kappa$ is a dimensionful scalar-field. Notably, one can quickly check that $\delta \Theta/\delta \kappa = 0 \Rightarrow \kappa = - R/\mathcal{L}_m$, as in Eq. (\ref{eq:f_Rkappa}). This \textit{Einstein-dilaton} form makes the additional degree-of-freedom of the theory manifest through the scalar-field $\kappa$. $\kappa$ simply is an additional gravitational field in addition to the metric field, which perturbations turn out to be even smaller than the perturbations of the metric field because of the \textit{intrinsic decoupling} mentionned above \cite{minazzoli:2013pr}.

Therefore, when gravity is entirely neglected, the path integral of Entangled Relativity can be approximated by
%Therefore, when one entirely neglects gravity, the path integral of Entangled Relativity can be approximated by
\begin{equation}
Z_{ER-QFT} \approx \int  \prod_i [D f_i] \exp \left(\frac{i}{\kappa \epsilon^2} \int d^4 x \mathcal{L}_m(f) \right). \label{eq:ERPIQFT}
\end{equation}
This path integral is particularly suitable for calculating phenomena at scales where gravity can be safely neglected—such as in colliders. To ensure the recovery of standard quantum field theory when gravity is disregarded, it follows that
%This path integral will notably be adequat for calculating any phenomenon at the scales at which gravity can safely be neglected---like in colliders for instance. Because one wants to recover standard quantum field theory when gravity is neglected, it means that
\begin{equation} \label{eq:equiv}
\kappa \epsilon^2 = c \hbar.
\end{equation}
This implies that $\hbar$ actually varies proportionally to $\kappa$ in general. In other words, $\hbar$ changes in a manner akin to a new gravitational field. It also establishes an explicit connection between the quantum and gravitational realms, indicating that $G \propto \hbar$, and that the weak gravity limit $G\rightarrow 0$ is also the classical limit $\hbar \rightarrow 0$.
%It implies that $\hbar$ actually varies proportionally to $\kappa$ in general. In other words, it means that $\hbar$ varies akin to a new gravitational field. It also gives an explicit connection between the quantum and the gravitational worlds, as it means that $G \propto \hbar$, and that the weak gravity limit $G\rightarrow 0$ also is the classical limit $\hbar \rightarrow 0$.

The variation of $\kappa$ is, however, expected to be very small throughout the observable universe. Indeed, it has been shown in \cite{arruga:2021pr} that the maximum amplitude variation for $\sqrt{1/\kappa}$ is only of the order of a few percents, from the center of the densest neutron stars to a distant observer. In the weak-field environment of the Solar System, the variation of $\hbar$ has been estimated to be $\delta \hbar / \hbar \approx 2.5 \times 10^{-12}$ between the surface of the Sun and a remote observer \cite{minazzoli:2024js}. It is important to note that these estimations are independent of any theoretical parameter, as there are no free theoretical parameters in the formulation of Entangled Relativity.
%The variation of $\kappa$ is nevertheless expected to be very small in all the observable universe. Indeed, it has been shown in \cite{arruga:2021pr} that the maximal amplitude variation for $\sqrt{1/\kappa}$ is of the order of a few percent only between the center of the densest neutron stars and a remote observer. In the weak-field environment of the Solar-System, the variation of $\hbar$ can been estimated to be $\delta \hbar / \hbar \approx  2.5 \times 10^{-12}$ between the surface of the Sun and a remote observer \cite{minazzoli:2024js}. Let us note that those estimations are independent of any theoretical parameter, since there is no free theoretical parameter in the formulation of Entangled Relativity.

%--------------------------------------------------------------
\subsection{The local (classical) vacuum limit:~}

Since $\mathcal{L}_m \neq \emptyset$ in Eq. (\ref{eq:ERPI}), Entangled Relativity inherently requires matter fields to permeate the entirety of spacetime. However, this does not necessarily mean that, classically, $\mathcal{L}_m$ cannot equal zero at some specific locations, though whether or not this can explicitly occur needs further investigation. Given that the scalar field $\kappa$ is defined as $\kappa = - R/\mathcal{L}_m$,
%Since $\mathcal{L}_m \neq \emptyset$ in Eq. (\ref{eq:ERPI}), Entangled Relativity enforces matter fields to permeat the whole spacetime by construction. However, it does not necessarely means that, classicaly, $\mathcal{L}_m$ cannot be equal to zero at some given position, although whether or not it can happen explicitely has to be investigated. But since the scalar-field $\kappa$ is given by $\kappa = - R/\mathcal{L}_m$
it may seem that $\kappa$ becomes singular as $\mathcal{L}_m \rightarrow 0$. However, this is not the case, as the behavior of $\kappa$ is dictated by the entire set of field equations, akin to how the constant behavior of $R/T$ in General Relativity is governed by its field equations in the limit $T \rightarrow 0$. Specifically, it can be verified that Eq. (\ref{eq:sceq}) remains well-behaved in the $(\mathcal{L}_m,T) \rightarrow 0$ limit. This implies that whenever $(\mathcal{L}_m,T) \rightarrow 0$, $R$ approaches zero at the same rate. This behavior is exemplified by the spherically charged black-hole solution in Entangled Relativity found in \cite{minazzoli:2021ej}. For this solution, both the Ricci scalar $R$ and the on-shell matter Lagrangian $\mathcal{L}_m \propto E^2$, where $E$ is the electric field, are proportional to the charge squared, while $T=0$. Consequently, the ratio between the two remains well-defined and constant as $\mathcal{L}_m \rightarrow 0$, as one should have anticipated directly from Eq. (\ref{eq:sceq}).

%the scalar-field seems to be singular at the limit $\mathcal{L}_m \rightarrow 0$. But in fact, it is not singular, because the behavior of $\kappa$ is dicated by the whole field equations, just as the constant behavior of $R/T$ in General Relativity  in the limit $T \rightarrow 0$ is imposed by the field equations of General Relativity. In particular, one can check that Eq. (\ref{eq:sceq}) is perfectly well-behaved in the ($\mathcal{L}_m,T) \rightarrow 0$ limit. This notably means that whenever $\mathcal{L}_m \rightarrow 0$, $R \rightarrow 0$ at the same rate. This is exemplified with the spherically charged black-hole of Entangled Relativity found in \cite{minazzoli:2021ej}. For that solution, both the Ricci scalar $R$ and the on-shell matter Lagrangian $\mathcal{L}_m \propto E^2$, where $E$ is the electric-field, are proportional to the charge squarred. It implies that the ratio between the two is perfectly well-defined and constant at the limit $\mathcal{L}_m \rightarrow 0$---as one should have anticipated directly from Eq. (\ref{eq:sceq}).

%--------------------------------------------------------------
\section{Conclusion}

In conclusion, Entangled Relativity completes the journey started by Einstein more than a century ago. It firmly embeds the principle of the relativity of inertia into its structure through a unique non-linear approach. This theory not only sticks closely to Einstein's original ideas but also creates a clear link between quantum mechanics and gravity, highlighted by the relationship $G \propto \hbar$. While we still need to explore how it behaves in the realm of quantum gravity, its framework already stands out by using fewer fundamental constants than General Relativity. At the same time, it matches the observed behavior of both General Relativity and standard quantum field theory in most cases. Entangled Relativity, therefore, isn't just yet another theoretical development; it's an actual novel general theory of relativity that might open new doors for understanding the universe.

%Entangled Relativity represents the culmination of the \textit{relativization} program initiated by Einstein almost a century ago, as it fundamentally incorporates the principle of relativity of inertia through its unique non-linear formulation. Moreover, it explicitly connects the quantum and gravitational realms with the derived relation $G \propto \hbar$, although its quantum gravity behavior still requires investigation.
%These characteristics alone make it a very promising theory. Additionally, its formulation is more parsimonious compared to General Relativity in terms of the required universal dimensionful constants. Remarkably, its phenomenology aligns with that of General Relativity and standard quantum field theory in most situations.

%Entangled Relativity is the finalization of the \textit{relativization} program start\-ed by Einstein almost a century ago, as it enforces the principle of relativity of inertia at its core, through its particular non-linear formulation. Besides that, it also explicitly bridges the quantum and gravitational realms with the (derived) relation $G \propto \hbar$, although its quantum gravity behavior remains to be investigated. 
%As if these were not already very promising properties, it also turns out that its formulation is more economical with respect to General Relativity in terms of universal dimensionful constants that are required in order to define it, while its phenomenology corresponds to the one of General Relativity and standatd quantum field theory in most situations. 

%\nocite{*}
\bibliographystyle{pepan}
\bibliography{ER_IWHEPFT}

\begin{thebibliography}{10}
\def\selectlanguageifdefined#1{
\expandafter\ifx\csname date#1\endcsname\relax
\else\selectlanguage{#1}\fi}
\providecommand*{\href}[2]{{\small #2}}
\providecommand*{\url}[1]{{\small #1}}
\providecommand*{\BibUrl}[1]{\url{#1}}
\providecommand{\BibAnnote}[1]{}
\providecommand*{\BibEmph}[1]{\emph{#1}}
\ProvideTextCommandDefault{\cyrdash}{\hbox to.8em{--\hss--}}
\providecommand*{\BibDash}{\ifdim\lastskip>0pt\unskip\nobreak\hskip.2em\fi
\cyrdash\hskip.2em\ignorespaces}

\bibitem{einstein:1918an}
\selectlanguageifdefined{english}
\BibEmph{{Einstein} A.} {Prinzipielles zur allgemeinen
  Relativit{\"a}tstheorie}~//
  \href{http://dx.doi.org/10.1002/andp.19183600402}{Annalen der Physik}.
  \BibDash
\newblock 1918. \BibDash
\newblock V. 360, no.~4. \BibDash
\newblock P.~241--244. \BibDash
\newblock Translation available at
  \url{https://einsteinpapers.press.princeton.edu/vol7-trans/49}.

\bibitem{hoefer:1995cf}
\selectlanguageifdefined{english}
\BibEmph{{Hoefer} C.} {Einstein's Formulations of Mach's Principle}~// Mach's
  Principle: From Newton's Bucket to Quantum Gravity~/ Ed.\ by\
  Julian~B.~{Barbour}, Herbert~{Pfister}. \BibDash
\newblock Boston University~: {Birkh\"aser}, 1995. \BibDash
\newblock P.~67.

\bibitem{pais:1982bk}
\selectlanguageifdefined{english}
\BibEmph{{Pais} A.} {Subtle is the Lord. The science and the life of Albert
  Einstein}. \BibDash
\newblock Oxford~: Oxford University Press, 1982.

\bibitem{einstein:1918sp}
\selectlanguageifdefined{english}
\BibEmph{{Einstein} A.} {Kritisches zu einer von Hrn. de Sitter gegebenen
  L{\"o}sung der Gravitationsgleichungen}~// Sitzungsberichte der K{\"o}niglich
  Preu{\ss}ischen Akademie der Wissenschaften (Berlin. \BibDash
\newblock 1918. \BibDash
\newblock P.~270--272. \BibDash
\newblock Translation available at
  \url{https://einsteinpapers.press.princeton.edu/vol7-trans/52}.

\bibitem{einstein:1913br}
\selectlanguageifdefined{english}
\BibEmph{Einstein A.} {On the Present State of the Problem of Gravitation}~//
  Phys. Z. \BibDash
\newblock 1913. \BibDash
\newblock V.~14. \BibDash
\newblock P.~1249--1262. \BibDash
\newblock Translation available at
  \url{https://einsteinpapers.press.princeton.edu/vol4-trans/210}.

\bibitem{norton:1995cf}
\selectlanguageifdefined{english}
\BibEmph{{Norton} J.D.} {Mach's Principle before Einstein}~// Mach's Principle:
  From Newton's Bucket to Quantum Gravity~/ Ed.\ by\ Julian~B.~{Barbour},
  Herbert~{Pfister}. \BibDash
\newblock 1995. \BibDash
\newblock P.~9.

\bibitem{barbour:1995cf}
\selectlanguageifdefined{english}
\BibEmph{{Barbour} J.} {Mach before Mach}~// Mach's Principle: From Newton's
  Bucket to Quantum Gravity~/ Ed.\ by\ Julian~B.~{Barbour}, Herbert~{Pfister}.
  \BibDash
\newblock Boston University~: {Birkh\"aser}, 1995. \BibDash
\newblock P.~6.

\bibitem{borzeszkowski:1995cf}
\selectlanguageifdefined{english}
\BibEmph{{Borzeszkowski} H., {Wahsner} R.} {Mach's criticism of Newton and
  Einstein's reading of Mach: The stimulating role of two misunderstandings}~//
  Mach's Principle: From Newton's Bucket to Quantum Gravity~/ Ed.\ by\
  Julian~B.~{Barbour}, Herbert~{Pfister}. \BibDash
\newblock Boston University~: {Birkh\"aser}, 1995. \BibDash
\newblock P.~58.

\bibitem{krauss:2023cf}
\selectlanguageifdefined{english}
\BibEmph{{Krauss} C.} {A Machian interpretation of the theory of relativity?
  Joseph Petzoldt's reading of Einstein}~// Philosophers and Einstein's
  Relativity~/ Ed.\ by\ Chiara~R.~{Krauss}, Luigi~{Laino}. \BibDash
\newblock Boston Studies in the Philosophy and History of Science~: {Springer},
  2023. \BibDash
\newblock P.~35.

\bibitem{einstein:1914sr}
\selectlanguageifdefined{english}
\BibEmph{Einstein A.} {Physikalische Grundlagen einer Gravitationstheorie}~//
  Naturforschende Gesellschaft in Zürich. Vierteljahrsschrift 58 (1914):
  284-290. \BibDash
\newblock 1914. \BibDash
\newblock V.~58. \BibDash
\newblock P.~284--290. \BibDash
\newblock Translation available at
  \url{https://einsteinpapers.press.princeton.edu/vol4-trans/204}.

\bibitem{einstein:1917co}
\selectlanguageifdefined{english}
\BibEmph{{Einstein} A.} {Kosmologische Betrachtungen zur allgemeinen
  Relativit{\"a}tstheorie}~// Sitzungsberichte der K{\"o}niglich
  Preu{\ss}ischen Akademie der Wissenschaften (Berlin. \BibDash
\newblock 1917. \BibDash
\newblock P.~142--152. \BibDash
\newblock Translation available at
  \url{https://einsteinpapers.press.princeton.edu/vol6-trans/433}.

\bibitem{einstein:1921bk}
\selectlanguageifdefined{english}
\BibEmph{{Einstein} A.} {The meaning of relativity}. \BibDash
\newblock Princeton~: Princeton University Press, 1921.

\bibitem{wilczek:2016bk}
\selectlanguageifdefined{english}
\BibEmph{{Wilczek} F.} {A Beautiful Question}. \BibDash
\newblock UK~: Penguin Random House, 2016.

\bibitem{MTW}
\selectlanguageifdefined{english}
\BibEmph{{Misner} C.W., {Thorne} K.S., {Wheeler} J.A.} {Gravitation}. \BibDash
\newblock 1973.

\bibitem{godel:1949rm}
\selectlanguageifdefined{english}
\BibEmph{{G{\"o}del} K.} {An Example of a New Type of Cosmological Solutions of
  Einstein's Field Equations of Gravitation}~//
  \href{http://dx.doi.org/10.1103/RevModPhys.21.447}{Reviews of Modern
  Physics}. \BibDash
\newblock 1949. \BibDash
\newblock V.~21, no.~3. \BibDash
\newblock P.~447--450.

\bibitem{loll:2008cq}
\selectlanguageifdefined{english}
\BibEmph{{Loll} R.} {The emergence of spacetime or quantum gravity on your
  desktop}~// \href{http://dx.doi.org/10.1088/0264-9381/25/11/114006}{Classical
  and Quantum Gravity}. \BibDash
\newblock 2008. \BibDash
\newblock V.~25, no.~11. \BibDash
\newblock P.~114006. \BibDash
\newblock arXiv:0711.0273~[gr-qc].

\bibitem{minazzoli:2022ar}
\selectlanguageifdefined{english}
\BibEmph{{Minazzoli} O.} {Quantum of action in entangled relativity}~//
  \href{http://dx.doi.org/10.48550/arXiv.2304.09482}{arXiv e-prints}. \BibDash
\newblock 2022. \BibDash
\newblock arXiv:2206.03824.

\bibitem{minazzoli:2023ar}
\selectlanguageifdefined{english}
\BibEmph{{Minazzoli} O.} {Standard quantum field theory from entangled
  relativity}~//
  \href{http://dx.doi.org/10.48550/arXiv.2304.09482}{Contribution to the 2023
  Gravitation session of the 57th Rencontres de Moriond}. \BibDash
\newblock 2023. \BibDash
\newblock P.~arXiv:2304.09482. \BibDash
\newblock arXiv:2304.09482.

\bibitem{ludwig:2015pl}
\selectlanguageifdefined{english}
\BibEmph{{Ludwig} H., {Minazzoli} O., {Capozziello} S.} {Merging matter and
  geometry in the same Lagrangian}~//
  \href{http://dx.doi.org/10.1016/j.physletb.2015.11.023}{Physics Letters B}.
  \BibDash
\newblock 2015. \BibDash
\newblock V. 751. \BibDash
\newblock P.~576--578. \BibDash
\newblock arXiv:1506.03278.

\bibitem{minazzoli:2013pr}
\selectlanguageifdefined{english}
\BibEmph{{Minazzoli} O., {Hees} A.} {Intrinsic Solar System decoupling of a
  scalar-tensor theory with a universal coupling between the scalar field and
  the matter Lagrangian}~//
  \href{http://dx.doi.org/10.1103/PhysRevD.88.041504}{Physical Review D}.
  \BibDash
\newblock 2013. \BibDash
\newblock V.~88, no.~4. \BibDash
\newblock P.~041504. \BibDash
\newblock arXiv:1308.2770~[gr-qc].

\bibitem{arruga:2021pr}
\selectlanguageifdefined{english}
\BibEmph{{Arruga} D., {Rousselle} O., {Minazzoli} O.} {Compact objects in
  entangled relativity}~//
  \href{http://dx.doi.org/10.1103/PhysRevD.103.024034}{Physical Review D}.
  \BibDash
\newblock 2021. \BibDash
\newblock V. 103, no.~2. \BibDash
\newblock P.~024034. \BibDash
\newblock arXiv:2011.14629.

\bibitem{minazzoli:2024js}
\selectlanguageifdefined{english}
\BibEmph{{Minazzoli} O.} {Solar System phenomenology of Entangled
  Relativity}~// Contribution to the 2023 proceedings of the "Journées
  Systèmes de Référence Spatio-temporels". \BibDash
\newblock 2024. \BibDash
\newblock https://journees2023.sciencesconf.org/.

\bibitem{minazzoli:2021ej}
\selectlanguageifdefined{english}
\BibEmph{{Minazzoli} O., {Santos} E.} {Charged black hole and radiating
  solutions in entangled relativity}~//
  \href{http://dx.doi.org/10.1140/epjc/s10052-021-09441-w}{European Physical
  Journal C}. \BibDash
\newblock 2021. \BibDash
\newblock V.~81, no.~7. \BibDash
\newblock P.~640. \BibDash
\newblock arXiv:2102.10541.

\end{thebibliography}

\end{document}